# Hybrid Team Tetris: A New Platform For Hybrid Multi-Agent, Multi-Human Teaming

Kaleb MCDOWELL [a,1], Nick WAYTOWICH [a], Javier GARCIA [a], Stephen GORDON [b], Bryce BARTLETT [b], and Jeremy GASTON [a]

[a] *Army Research Laboratory*
[b] *DCS Corporation*

**Abstract.** Metcalfe et al (1) argue that the greatest potential for human-AI partnerships lies in their application to highly complex problem spaces. Herein, we discuss three different forms of hybrid team intelligence and posit that across all three forms, the hybridization of man and machine intelligence can be effective under the right conditions. We foresee two significant research and development (R&D) challenges underlying the creation of effective hybrid intelligence. First, rapid advances in machine intelligence and/or fundamental changes in human behaviors or capabilities over time can outpace R&D. Second, the future conditions under which hybrid intelligence will operate are unknown, but unlikely to be the same as the conditions of today. Overcoming both of these challenges requires a deep understanding of multiple human-centric and machine-centric disciplines that creates a large barrier to entry into the field. Herein, we outline an open, shareable research platform that creates a form of hybrid team intelligence that functions under representative future conditions. The intent for the platform is to facilitate new forms of hybrid intelligence research allowing individuals with human-centric or machine-centric backgrounds to rapidly enter the field and initiate research. Our hope is that through open, community research on the platform, state-of-the-art advances in human and machine intelligence can quickly be communicated across what are currently different R&D communities and allow hybrid team intelligence research to stay at the forefront of scientific advancement.

**Keywords.** Hybrid Intelligence, Team Intelligence, Human-Machine Teaming, Research Platform, Research Environment

## 1. Introduction

Increased hybridization of intelligence between humans and artificial capabilities is envisioned arising from decades of rapid, pervasive, and disruptive technological evolution (see 2,3). Herein, we focus on hybrid team intelligence, which we define as the ability to learn or use skilled reasoning (intelligence) that arises from multi-agent, multi-human systems (teams) where there is a deliberate blending of the processing between at least two disparate disciplines, cultures, or entities (hybridization). In this definition, any specific entity does not need to be intelligent. Rather, the mixed human-agent system overall must present the ability to learn or use skilled reasoning.

Numerous scientific and commercial breakthroughs point to disruptive technological evolution already forming the basis for hybrid team intelligence. We give examples of three forms of human-technology interaction: technology as a tool, technology as a human decision enhancer, and human-guided artificial learning. Initial disruptions came in the form of humans using tools that solve complex problems. The creation of consumer Global Positioning System (GPS) navigation devices pioneered by Garmin in the 1990s exemplifies how complex problem-solving tool use can reduce, but not eliminate, human cognitive demand associated with specifics aspects of navigation (see 4). Extending this form of hybrid intelligence to teams, the development of weather forecasting models has led to a dramatically altered approach to weather forecasting involving numerous complex models with numerous human users to generate hybrid forecasts with less errors than models or humans alone can produce (e.g., see National Hurricane Center). Technological advances have gone beyond human use of tools to relationships where technologies integrate, modify, and extend human decisions. The creation of Foldit for predicting protein structures provides an example of this form of relationship by merging crowdsourcing with intelligent technologies for ideation and discovery (5). The creation of cortically coupled computer vision technologies provides another example of integrated human inputs by combining brain-computer interfaces (BCI) along with computer vision algorithms to achieve extremely fast object detection at human-level performance, allowing for the triage of large image databases (e.g., 6,7). Most recently, human-guided machine learning techniques have evolved that shift humans from the roles of actors to that of technology teachers. Warnell et al (8) demonstrate this form of human-machine interaction via having a human trainer teach a deep learning agent to play Atari games using evaluative human feedback called TAMER (Teaching Agents Manually through Evaluative Reinforcement). Using Deep TAMER, Atari games can be taught in less than 15 mins; a task that has proven difficult for state-of-the-art reinforcement learning methods. At a group level, this form of interaction is also observed in the commercial sector in selector systems that use inputs, such as thumbs up or thumbs down, to prioritize playlists, advertisements, etc. A key finding across all three forms of human-technology relationship described herein is the demonstration that hybrid intelligence *can* outperform a machine-only or human-only approaches under the right conditions.

Focusing on future human-AI ecosystems, Metcalfe et al. (1) argue that the greatest potential for human-AI partnerships lies in their application to highly complicated problem spaces. Herein, we embrace a vision of hybrid team intelligence applied in such complex problem spaces. Our first aim is to articulate three factors driving the types of problem spaces hybrid team intelligence may solve. Our second aim is to present a vision for enabling community research with a novel hybrid team intelligence research platform to enable that vision.

## 2. A Complex Problem Space for the Future Hybrid Team Intelligence

The rapid evolution of machine intelligence presents fundamental challenges to the research and development community. First, advances in machine intelligence can outpace the rate at which the community develops its understanding of relationship between intelligent technologies and humans; that is, the interactions or relationships between humans and technology studied for a particular state-of-the-art technology can quickly be rendered irrelevant by either the significant advancement of the technology or the rise of a superior machine intelligence. Second, the advances in intelligent technologies are correlated with changes in human behaviors and potentially human capabilities; for example, there are numerous negative short- and long-term effects on human spatial reasoning resulting from GPS device use (4,9) as well as changes in human roles (human knowledge, skills, and behavior requirements) in the aforementioned weather forecasting and human-guided machine learning examples. This correlation effectively points to the potential for human-machine intelligence research to be studying the "wrong people," which can directly impact overall performance as well as underlying issues such as system usability and trust (see 10). We posit that overcoming these challenges will require the combined expertise of multidisciplinary teams (of humans and technology) with an emphasis on combining human-centric and technology-centric expertise.

Consistent with the literature on multidisciplinary teaming (see 1, 11, 12), we have observed that specialized perspectives, communications, priorities, and assumptions of human-centric versus intelligent technology-centric researchers and developers can make it quite difficult for teams to function effectively together. Our approach to overcoming the challenges of developing high performing mixed human- and machine intelligence-centric research and development teams starts with attempting to create a shared vision for the future. While numerous future environments are plausible, we constrain our vision through three "driving factors" that we extract from current technological and sociotechnical trends:

- **Advanced Intelligence:** Machine intelligence technologies that sense, perceive, reason, and/or learn are expected to continue to become increasingly capable, more pervasive in society, and potentially more "alien" to humans. These technologies will adapt to and learn from a variety of sources including humans, other technologies, as well as the environment. While not as dramatic, human intelligence and its application is also expected to become refined with the continued development of cognitive enhancement technologies, training approaches, and continued exposure to advanced sociotechnical ecosystems. *Sample Challenge: Create future human-technology interactions for human and machine intelligence that does not yet exist.*
- **"Superhuman" Capabilities:** The continued evolution and growth of interconnected sociotechnical ecosystems combined with more capable data processing and interpretation capabilities will continue to create conditions where decisions are made faster and/or on more data than humans can effectively process through traditional means. *Sample Challenge: Create future human-technology interactions that allow for humans to influence "superhuman" decisions and actions without limiting overall performance.*
- **Rapid Technological Change:** Continued decreases in the time from ideation to fielding of new technologies combined with the fact that these technologies have become increasingly pervasive in society has created an environment of constant and rapid technological change. *Sample Challenge: Enable humans and machines to have effective and stable in-field adaptations to predicted and unforeseen changes in the sociotechnical environment.*

Combined, these three interrelated "driving factors" constrain the complex problem space in which we expect hybrid team intelligence to be critical. However, we have observed that this level of constraint is still insufficient to bring together high performing mixed human- and intelligent technology-centric research and development teams. To help multidisciplinary teams further create a shared vision, we have attempted to instantiate this future problem space in the simplified, *toy* research platform of Hybrid Team Tetris.

**3. Hybrid Team Tetris**

The primary aim of Hybrid Team Tetris is to create hybrid team intelligence within the envisioned complex problem space described above *and* to do so in a manner that makes the platform easily accessible to a broad community of human-centric and machine intelligence-centric researchers and developers. Hybrid Team Tetris is a functional multi-agent, multi-human open-source research platform that allows for users with different knowledge, skills and behaviors to experience key elements of hybrid team intelligence as outlined below. The intent of the platform is to enable productive multidisciplinary discussions through the creation of critical technical and research challenges and ongoing community contributions. While this paper provides only a brief introduction to Hybrid Team Tetris, downloadable software, as well as, a description of an upcoming open community Hybrid Team Tetris-based hack-a-thon (see 'Human System Adaptation Paradigm Challenge 2022') can be found on the GitHub website as of Jun 17, 2022 (https://github.com/DCSHGAI/HGAITetris).

*3.1. Why Tetris*

While numerous experimental and gaming paradigms could serve as the basis for our community platform (see 'Human System Adaptation Paradigm Challenge 2022' for opportunities to present alternative platforms), the choice to use Tetris as a basis for the platform arose from three primary reasons:

- Tetris has been successfully deployed as a basic form of hybrid intelligence in the past (13, 14)
- Tetris is well-defined and has been used as a platform for both human-centric and machine intelligence-centric research and development for over two decades (15-17; Current Tetris AI Record set by G Cannon in 2021)
- Tetris offers experimental flexibility and accessibility (for multiple versions and configurations see 18-20).

In addition, Tetris is one the of the best-selling games of all time (>100M copies sold), although we acknowledge that Hybrid Team Tetris has very different game play than that originally designed by Alexey Pazhitnov in 1984.

*3.2. Basic Multi-Agent, Multi-Player Gameplay*

At its core, Hybrid Team Tetris uses a human-guided machine learning approach similar to Knox & Stone's (13), but updated with Deep TAMER (see 8). That is, as the machine agent drops a "tetrimino" into a row, a human has an option to press or not press "ENTER" to indicate what they view as a "positive" move. The Deep TAMER uses the reward signal to relate the tetrimino shape and dropped position to the state of the game board and update its game play. This basic mechanism allows humans, who may have no subject matter expertise in intelligent technologies, to train agents to play the game.

To extend the game to multi-agent, multi-player gameplay, three mechanisms are employed. First, a single player is presented with multiple games simultaneously. That is, the player can use keystrokes ("1-2") to toggle between games and then, press "ENTER" to indicate a positive move in the selected game. This allows the player to develop independent agents. Second, a multi-player mode allows additional players to also play simultaneous games and develop other independent agents, using the aforementioned mechanism. Third, to develop interactions across the team, additional games are played by machine agents that learn from all or a sub-set of the player directed games. This creates agent learning that is dependent on team decisions. A functional version of the game with 2 players each guiding 2 agents with one dependent agent learning from the 4 player-guided games is depicted in figure 1 and is located on the GitHub website (see Base platform).

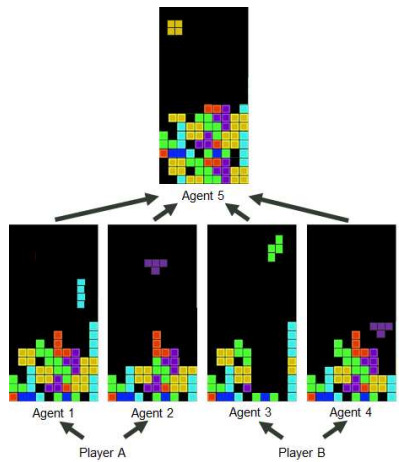

**Figure 1.** Basic Hybrid Team Tetris gameplay. Human Player A guides machine Agents 1 & 2. Human Player B guides machine Agents 3 & 4. Machine Agent 5 learns from machine Agents 1, 2, 3, and 4.

The number of total numbers of players, agents, and the network interconnectivity of players and agents can be manipulated to study issues that range from team collaboration to distributed learning.

*3.3. Advanced Intelligence Gameplay*

A critical challenge for studying future human-technology hybridization is to create a research environment that is representative of the advances in human and machine intelligence that do not yet exist. Hybrid Team Tetris approaches this challenge through the use of "Hidden Rules" that are only initially presented to a single or a sub-set players or agents. For example, if the "Hidden Rule" is that anytime a row is cleared with >3 yellow squares, a "favorable" tetrimino is presented as the next piece (see Spiel et al 2017 for adapting Tetris difficulty via adjusting the next available Tetrimino) and the agent receives a "positive" reward even if the human-player does not actively press "ENTER". Alternatively, whenever the row with >3 yellow squares is cleared, a "10*bonus score" could be awarded immediately on screen if the player is being presented the rule or as an end of game bonus if the agent is being presented the rule. A functional version of the basic game with the later "Hidden Rule" is located on the GitHub website (See Advanced_Intelligence platform).

*3.4. "Superhuman" Gameplay*

A second critical challenge for human-technology hybridization is effectively allowing humans to influence "superhuman" decisions and actions. The original game of Tetris naturally generates this challenge by ramping up the speed at which Tetriminos are dropped. Hybrid Team Tetris ramps up this challenge by adding multiple simultaneous games per player (up to 10 selected via keystrokes "0-9") in addition to speed modifications. Extending the Knox & Stone (2009) version of human-guided machine learning Tetris allows players to overcome this challenge by granting access to discrete moments in time rather than continuously making decisions. A functional version of the basic game with "superhuman" challenges is located on the GitHub website (See Superhuman platform).

*3.5. Rapid Technological Change Gameplay*

A third critical challenge for human-technology hybridization is to enable humans and machines to have effective and stable in-field adaptations to predicted and unforeseen changes in the sociotechnical environment. Hybrid Team Tetris allows for this challenge by introducing in-game changes that require substantive changes in strategy to maintain gameplay and serves as a proxy for rapid technological change. For example, the version on GitHub combines the introduction of novel Tetrimino pieces and alters the rules awarding points as game time elapses (see Rapid_Technological_Change platform).

*3.6. Overall Gameplay*

While all the above examples of Hybrid Team Tetris are functional, our goal is to demonstrate a hybrid intelligence suitable research platform that represents our envisioned future complex problem space, which combines advanced intelligence, "superhuman" capabilities, *and* rapid technological change. A functional version of the combined platform is also available on GitHub (See Integrated platform).

**4. Conclusions**

Hybrid Team Tetris is an open, shareable research platform that aims to lower the bar for entry into hybrid intelligence by allowing individuals to rapidly experience hybrid teams and lowers the multidisciplinary knowledge requirement to initiate research. Moreover, community contribution on this open-source platform allows for increasing sophistication of both human-centric and machine-centric interactions. This sophistication across R&D communities allows for effective cross-community state-of-the-art collaboration and advancement of hybrid intelligence principles and insights.